\RequirePackage{ifpdf}
\documentclass{PoS}

 \usepackage{graphicx}
\ifpdf
\usepackage{epstopdf}
\fi
\usepackage[multidot]{grffile}

\usepackage{fixltx2e}

 \usepackage{graphics}
 \usepackage{caption}
 \usepackage{epsfig}
 \usepackage{longtable}
 \usepackage{latexsym}
 \usepackage{amsmath}
 \usepackage{amsthm}
 \usepackage{amsfonts}
 \usepackage{amssymb}
 \usepackage{dsfont}
 \usepackage{bm}
 \usepackage{subfigure}
 \MakeRobust{\eqref}
\graphicspath{{figs/}}

\newcommand{\figs}{figs/}

\renewcommand{\>}{\rangle}
\renewcommand{\dag}{^\dagger}

\newcommand{\<}{\langle}

\newcommand{\be}{\begin{equation}}
\newcommand{\ee}{\end{equation}}
\newcommand{\bi}{\begin{itemize}}
\newcommand{\ei}{\end{itemize}}
\newcommand{\bey}{\begin{eqnarray}}
\newcommand{\eey}{\end{eqnarray}}

\newcommand{\LambdaR}{\ensuremath{\Lambda_{\rm R}}}
\newcommand{\mR}{\ensuremath{m_{\rm R}}}
\newcommand{\rhoR}{\ensuremath{\tilde\rho_{\rm R}}}
\newcommand{\msbar}{{\rm \overline{MS\kern-0.05em}\kern0.05em}}

\title{Chiral condensate in $n_f=2$ QCD from the Banks--Casher relation}
\ShortTitle{Chiral condensate in $n_f=2$ QCD from the Banks--Casher relation}

\author{\speaker{Georg P.~Engel}
\\
Dipartimento di Fisica, Universit\`a Milano-Bicocca, \\
and INFN, Sezione di Milano-Bicocca, \\
Piazza della Scienza 3, 20126 Milano, Italy \\
E-mail: \email{georg.engel@mib.infn.it}
}
\abstract{
Exploiting the Banks-Casher relation, we present a direct determination of the chiral condensate in two-flavor QCD, computing the mode number of the $\mathcal{O}(a)$-improved Wilson-Dirac operator below various cutoffs. We make use of CLS-configurations with three different lattice spacings in the range of 0.05-0.08 fm and pion masses down to 190 MeV. Our data indicate a non-zero density of eigenmodes near the origin and hence points to spontaneous chiral symmetry breaking. We extrapolate our results to the continuum and chiral limit to give a result for the chiral condensate.
}

\FullConference{The XXXII International Symposium on Lattice Field Theory, Lattice 2014\\
June 23 - 28, 2014\\
Columbia University, New York, USA}

\begin{document}

\section{Introduction}
\label{sec:intro}
\noindent
The chiral condensate, defined as expectation value of a quark-antiquark pair,
\be
\Sigma \equiv -\frac{1}{2}\<\bar\psi\psi\> \;,
\ee
plays a central r\^{o}le in QCD. 
It provides an order parameter for chiral symmetry breaking, a leading-order low-energy constant of Chiral Perturbation Theory (ChPT), and it naturally appears in the Operator Product Expansion. 
Recent related lattice QCD results are collected in the FLAG review \cite{Aoki:2013ldr}.
The present work discusses a determination of $\Sigma$, exploiting the Banks-Casher relation \cite{Banks:1979yr}, 
\bey
\Sigma 						&=& \pi \lim_{\lambda \to 0} \lim_{m \to 0} \lim_{V \to \infty} \rho(\lambda,m) \;,\\
\text{with} \qquad \rho(\lambda,m) 	&=& \frac{1}{V}\sum_{k=1}^{\infty} \< \delta(\lambda-\lambda_k) \> \;, 
\eey
where $m$ is the current quark mass, $i\lambda_k$ are the eigenvalues of the massless Dirac operator and $V$ is the four-volume. 
The spectral density $\rho$ is renormalizable and can be computed on the lattice numerically \cite{Giusti:2008vb}.
In lattice QCD with Wilson-type quarks, it turns out to be convenient to consider the mode number $\nu(\Lambda,m)$ of the massive hermitian operator $D_m\dag D_m$ with eigenvalues $\alpha\leq M=\sqrt{\Lambda^2+m^2}$, which is renormalization-group invariant, 
\bey
\nu(\Lambda,m)			&=& V \int_{-\Lambda}^{\Lambda} d\lambda \rho(\lambda,m) \\
\nu_R(\Lambda_R,m_R)		&=& \nu(\Lambda,m)\;.
\eey
The method was shown to work in Ref.~\cite{Giusti:2008vb} and applied to twisted--mass fermions in Ref.~\cite{Cichy:2013gja}.  
We define the effective spectral density, 
\be
\rhoR	= \frac{\pi}{2 V} \frac{\nu_{2,\rm R}-\nu_{1,\rm R}}{\Lambda_{2,\rm R} - \Lambda_{1,\rm R}} \;,
\ee
which agrees with $\Sigma$ after taking the appropriate order of limits as in the Banks-Casher relation. 
Note that any threshold effects are removed from $\rhoR$ as long as all $\Lambda_{i, \rm R}$ are chosen large enough. 
Preliminary results have been presented in Ref.~\cite{Engel:2013rwa}, the main physics results are published in Ref.~\cite{Engel:2014cka}, while for a detailed discussion we refer to Ref.~\cite{Engel:2014xx}.

\section{Chiral Perturbation Theory}
\label{sec:chpt}
\noindent
At next-to-leading-order (NLO), continuum ChPT predicts for $n_f=2$ QCD \cite{Giusti:2008vb},
\be\label{eq:chpt}
\hspace{-3mm}
\rhoR^{\rm NLO}(\Lambda_{\rm 1,R},\Lambda_{\rm 2,R},\mR)  = \Sigma \Big\{1 + \frac{\mR \Sigma}{(4\pi)^2 F^4}
									 \Big[ 3\, \bar l_6 + 1 - \ln(2) - 3 \ln\Big(\frac{\Sigma \mR}{F^2 \mu^2}\Big)
									+ \tilde g_\nu\left(\frac{\Lambda_{\rm 1,R}}{\mR},\frac{\Lambda_{\rm 2,R}}
									 {\mR}\right)\Big]    \Big\} 	\;,
\ee
where $\tilde g_{\nu}(x_1,x_2)$, explicitly given in Ref.~\cite{Engel:2013rwa}, appears to be a mild function in the considered range of parameters \cite{Engel:2014xx}. 
$F$ is the pseudo-scalar decay constant in the chiral limit, $\bar l_6$ an NLO low-energy constant (LEC) and $\mu$ is a fixed scale.
It is noteworthy that there are no chiral logarithms at fixed $\LambdaR$, 
that ${\rhoR}^{\rm NLO}$ is a decreasing function of $\LambdaR=(\Lambda_{\rm 1,R}+\Lambda_{\rm 2,R})/2$ for any finite quark mass, and 
also that in the chiral limit all NLO-corrections vanish in the two-flavor theory \cite{Giusti:2008vb,Smilga:1993in}. 

NLO Wilson-ChPT, considering $\mathcal{O}(a)$ improvement and the generically-small-quark-mass regime (GSM) \cite{Necco:2011vx}, gives an additional term of the form $\mR/(\Lambda_{\rm 1,R}\Lambda_{\rm 2,R})$ added to Eq.~\eqref{eq:chpt}. The sign of this term was argued to be positive   \cite{Hansen:2011kk,Splittorff:2012gp}, 
which implies that ${\rhoR}^{\rm NLO,lat}$ is a decreasing function of $\LambdaR$ also at finite lattice spacing (in the GSM-regime). 
We remark that those NLO discretization effects, and any $\LambdaR$-dependence, are still absent in the chiral limit. 
The formalism of ChPT can be used also to address finite-volume effects, which increase towards light $\LambdaR$.

\section{Details of the simulation}
\label{sec:simdet}
\noindent
\begin{table}
\small
\begin{center}
\setlength{\tabcolsep}{.10pc}
\begin{tabular}{@{\extracolsep{1mm}}ccccccccccc}
\hline
\hline
id  & $L/a$  &$m_\pi$ [MeV]&$m_\pi L$&$a$ [fm]	& $R\tau_{\rm exp}$	& $R\tau_{\rm int}(m_{\pi})$ & $R\tau_{\rm int}(\nu)$ & $Rn_{\rm it}(\nu)$	&$N_{\rm cnfg}$ \\
\hline
 A3   & $32$& $496(6)$  & $6.0$  & 0.0749(8)	& 40	& 7	& 					& 47.36		& 55\\     
A4   &           & $386(5)$  &  $4.7$  &			& 	& 5	&				& 53.28		& 55\\
A5   &            &$333(5)$  & $4.0$   &			& 	& 5	& 3				& 36.00		& 55\\
B6   & $48$   & 283(4)    & 5.2   	&  			 & 	& 6	& 				& 24.00		& 50\\
\hline
E5  &$32$   &$ 440(5)$  & $4.7$ &  0.0652(6)  & 55	& 9	& 6					& 35.52		& 92\\
F6  &$48$     &$ 314(3)$  & $5.0$ &  			&	& 8	&				& 29.60		& 50\\
F7  &           &$ 268(3)$  & $4.3$ &    			&	& 7	&				& 26.64		& 50 \\
G8  &$64$     &  193(2)       &   4.1 		&   	&	& 8	&				& 24-48		& 50 \\
\hline
N5 & 48  & 443(4)  & 5.2 &  0.0483(4)  		& 100	& 30		& 11			& 28.16	& 	60 \\
N6 &       & 342(3)  & 4.0 &  				& 		& 10		& 			& 128.0	& 	60 \\
O7 & 64  & 269(3)  & 4.2 &  				& 		& 15		& 			& 76.00		& 	50 \\
\hline
\hline
\end{tabular}
\caption{
Parameters of the simulation. $L$ is the linear spatial extent of the lattice, $a$ the lattice spacing \cite{Fritzsch:2012wq}, $m_\pi$ the pion mass, $R$ the ratio of active links in DD-HMC \cite{Luscher:2003qa} ($R=1$ in MP-HMC \cite{Marinkovic:2010eg}).  
$\tau_{\rm exp}$ and $\tau_{\rm int}$ denote the exponential and integrated autocorrelation time, resp., given in units of molecular dynamics, 
$n_{\rm it}$ the separation of configurations between subsequent measurements and $N_{\rm cnfg}$ the number of configurations on which $\nu$ is measured. 
}
\vspace{-3mm}
\label{tab:simdet}
\end{center}
\end{table}
We measure the mode number on configurations with two flavors of $\mathcal{O}(a)$-improved Wilson quarks, provided by the CLS-collaboration. 
The most relevant details for the present study are depicted in Tab.~\ref{tab:simdet}, further information is detailed in Refs.~\cite{Fritzsch:2012wq,Marinkovic:2011pa}. 
Finite-size effects and autocorrelations are under control for all measurements. 
The mode number is computed for nine values of $\LambdaR$ in the range 20-120 MeV with a statistical precision of a few percent on all ensembles. 
Rational polynomials are used to approximate the the spectral projector $\mathbb P_M$ to the low modes of the Dirac operator. 
Its expectation value is then evaluated stochastically with pseudo-fermion fields $\eta_k$, 
\be
\nu 	= \frac{1}{N}\sum_{k=1}^N \< \left( \eta_k, \mathbb P_M \eta_k \right)\> \;, \qquad\qquad M=\sqrt{\Lambda^2+m^2}\;.  
\ee

\section{Results}
\label{sec:results}
\begin{figure}[!t]
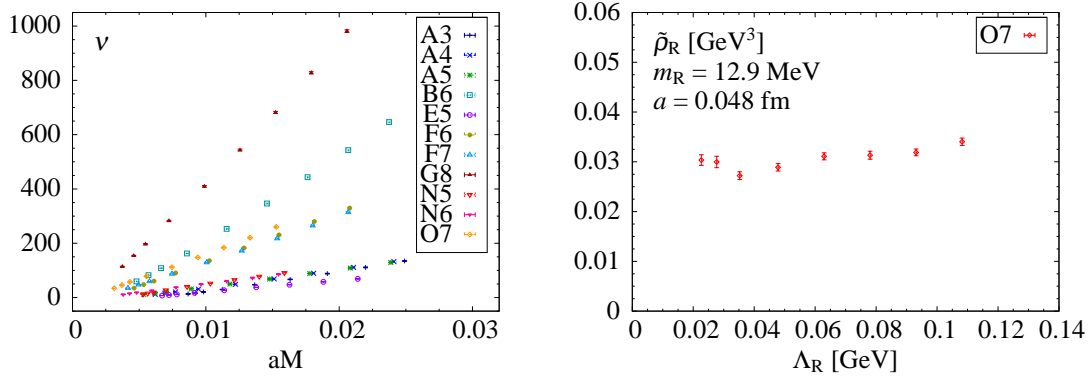

\begin{center}
{\LARGE\resizebox{!}{0.34\textwidth}{\input{\figs all-nu}}} 
{\LARGE\resizebox{!}{0.34\textwidth}{\input{\figs all-st-reduced-to-O7}}}
\caption{First look at the numerical data. 
Left: Mode number $\nu$ for all ensembles vs.~the bare dimensionless cutoff $aM$. Note the approximate linearity and the high number of modes achieved for small quark masses. 
Right: Effective spectral density $\rhoR$ vs.~the cutoff $\LambdaR$ for the ensemble with the lightest quark mass at the finest lattice spacing (O7). Note the non-zero flat behavior, which can be interpreted as a first hint for dynamical chiral symmetry breaking.} 
\label{fig:all-nu}
\end{center}
\end{figure}

\noindent
Fig.~\ref{fig:all-nu}, left, shows the results for the mode number for all ensembles. 
It exhibits a roughly linear dependence on $aM$ in all cases up to approximately 100-150 MeV. 
A phenomenological low-order polynomial fit indicates that in the considered range roughly 90\% of $\nu$ is given by the linear term. 
The effective spectral density $\rhoR$ shows a non-zero and flat behavior in $\LambdaR$ at fine lattice spacings and light quark masses. 
As an example, the results of ensemble O7 are shown in Fig.~\ref{fig:all-nu}, right. 

\begin{figure}[!t]
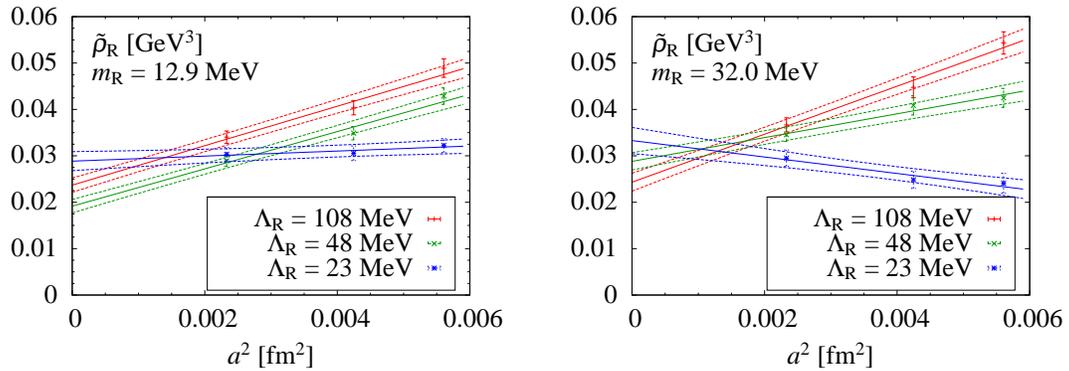

\begin{center}
{\LARGE\resizebox{!}{0.34\textwidth}{\input{\figs oh1014-m0,0129-lammulti-st-vs-a2}}}
{\LARGE\resizebox{!}{0.34\textwidth}{\input{\figs oh1014-m0,0320-lammulti-st-vs-a2}}}\\
\caption{Continuum extrapolation of $\rhoR$, performed individually for each pair $(\LambdaR,\mR)$. 
Shown for three values of $\LambdaR$ covering the entire range, and for the lightest (left) and the heaviest reference quark mass (right). Note that the data agrees well with the linear $a^2$-dependence expected in the $\mathcal{O}(a)$-improved theory. The discretization effects exhibit a non-trivial dependence on $(\LambdaR,\mR)$, but appear mild at the lightest point.} 
\label{fig:cont-extrapol}
\end{center}
\end{figure}

To extrapolate to the continuum and chiral limit, some analytic guidance is needed.  
In this respect, first studies indicated that higher-order effects of ChPT are apparent in the data and correspondingly the functional form at finite lattice spacing is not entirely clear in the considered range of parameters \cite{Engel:2013rwa}. 
For this reason, we attempt to build a clean fitting strategy where different effects can be distinguished clearly. 
Such a strategy is based on performing first the continuum limit, and only then 
removing the small corrections stemming from finite $\mR$ and $\LambdaR$. 
To do so, we interpolate $\rhoR$ to three values of the quark mass ($\mR= 12.9, 20.9, 32.0$ MeV) at each lattice spacing. 
A continuum extrapolation is then performed separately for each pair of $(\LambdaR,\mR)$, 
examples of which are shown in Fig.~\ref{fig:cont-extrapol}.
The linear $a^2$-dependence, expected from Symanzik effective theory for the $\mathcal{O}(a)$-improved theory, is respected well by the data. 
It is noteworthy that the discretization effects exhibit a non-trivial dependence on $(\LambdaR,\mR)$, but appear  fairly mild at the lightest points. 

\begin{figure}[!t]
\begin{center}
{\LARGE\resizebox{!}{0.4\textwidth}{\input{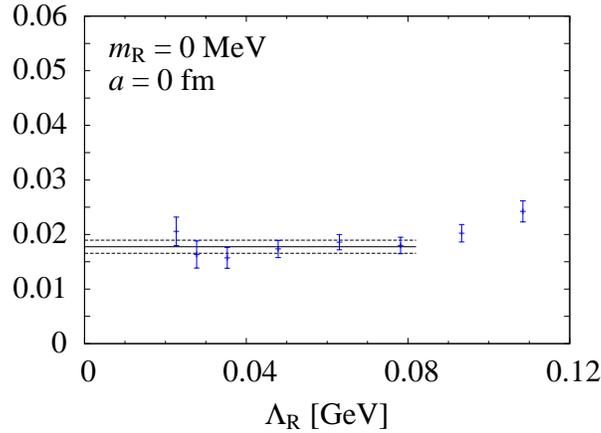}}}\\
\caption{The effective spectral density $\rhoR$ in the continuum and chiral limit.
The flat and non-zero behavior observed for $\Lambda\leq80$ MeV is consistent with NLO ChPT, 
a plateau fit in this range yields a prediction for the chiral condensate given in the text.} 
\label{fig:cont-chiral-lim}
\end{center}
\end{figure}

As a result of the extrapolation, we obtain $\rhoR$ in the continuum, where its non-zero values at light $(\LambdaR,\mR)$ point to dynamical chiral symmetry breaking. 
This motivates to use ChPT to remove the remaining small corrections. 
We consider a fit function based on generalized NLO ChPT,
\be
\rhoR = c_0(\LambdaR) + c_1\mR + c_2 \mR \left[ \tilde g_\nu\left(\frac{\Lambda_{\rm 1,R}}{\mR},\frac{\Lambda_{\rm 2,R}}{\mR}\right) -3\ln\left(\frac{\mR}{\mu}\right)\right]\;,
\ee
where $c_0(\LambdaR)=\Sigma={\rm const.}$ at NLO.
The continuum data is described well by this ansatz (the correlated fit gives $\chi^2/$dof=16.4/14), 
the extrapolation is of the order of the statistical error, 
and we obtain the results for $c_0(\LambdaR)$ shown in Fig.~\ref{fig:cont-chiral-lim}. 
The plateau-like behavior for $\LambdaR\leq80$ MeV indicates the NLO range, and a corresponding fit gives $\Sigma^{1/3}=261(6)$ MeV in the $\msbar$ scheme at 2 GeV.

To substantiate the result, we consider a second strategy to extract the chiral condensate from the data on the effective spectral density. 
After the separate treatment of different effects in the first strategy, we now attempt to perform a combined fit in $(\LambdaR,\mR,a)$ at once. 
The advantages are that this approach does not require an interpolation in $\mR$ but includes all data, and furthermore needs fewer fit parameters compared to the first strategy. 
However, ChPT is used from the start and the discretization effects have to be modelled. 
We assume a linear dependence in $a^2$ and $\mR$, but still allow for an arbitrary $\LambdaR$-dependence,  inspired by Symanzik effective theory and the chiral power expansion. 
It is worth noting that the model complies with the results of the first strategy and includes NLO Wilson-ChPT \cite{Necco:2011vx} as a special case. 
We find that the fit describes the data well and that the results agree very well with the ones of the first strategy. 
Having abundant degrees of freedom in the fit, we use the second strategy to estimate the systematic uncertainty  of the final result by performing various different fits. 
An upward shift is found when neglecting data at coarse lattices, while a downward shift is found when including higher-order terms $\mathcal{O}(\LambdaR^2,\mR^2)$ in the fit. 

\begin{figure}[!t]
\begin{center}
\includegraphics[width=100mm,clip]{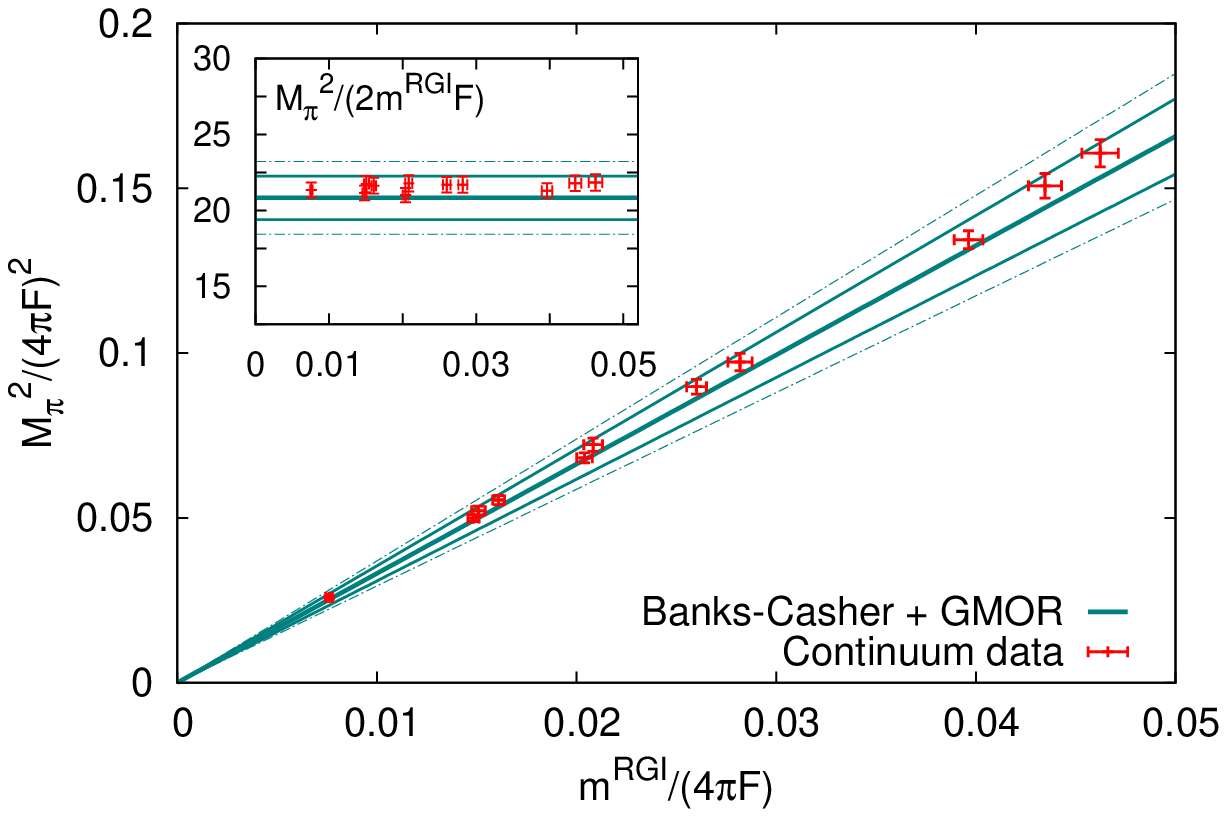}               
\caption{Consistency of the determined chiral condensate with the quark mass dependence of the pion mass as expected from the GMOR-relation. 
The pion mass squared $M_\pi^2$ is shown vs.~the renormalizion-group-independent (RGI) quark mass, normalized to $4\pi F$ ($\approx1$ GeV), where $F$ is the pseudo-scalar decay constant in the chiral limit (taken from \cite{Engel:2014cka}). 
The direct measurements (red symbols) are extrapolated to the continuum as described in Ref.~\cite{Engel:2014cka}, while the (central) solid line represents the GMOR contribution to the pion mass squared, computed by taking the direct determination of the chiral condensate through the spectral density. 
The thinner solid lines denote the statistical error, the dotted-dashed ones the sum of statistical and systematic one.
} 
\label{fig:gmor}
\end{center}
\end{figure}

\section{Conclusions}
\label{sec:conclusions }
\noindent
We presented a determination of the chiral condensate based on an extensive discussion of the spectral density of the hermitean Wilson Dirac operator. 
Our final result is 
\be
[\Sigma^{\msbar}(2~{\rm GeV})]^{1/3} = 261(6)(8)~{\rm MeV} \;,
\ee
where the first error is statistical and the second one systematic.
As a consistency test of dynamical symmetry breaking and ChPT, we consider its NLO prediction for the quark mass dependence of the pion mass. 
The latter is known as GMOR-relation and we show its prediction based on our measurement of the chiral condensate together with the direct measurements of the quark and pion masses in Fig.~\ref{fig:gmor}. 
The relation appears to be fulfilled to very good precision in the range considered.

\acknowledgments
\noindent
Simulations have been performed on BlueGene/Q at CINECA 
(CINECA-INFN agreement, ISCRA project IsB08\_Condnf2), on HLRN, on JUROPA/JUQUEEN at 
J\"ulich JSC, on PAX at DESY, Zeuthen, and on Wilson at Milano--Bicocca. 
We thank these institutions for the computer resources and the technical 
support. We are grateful to our colleagues within the CLS initiative for 
sharing the ensembles of gauge configurations. We acknowledge 
partial support by the MIUR-PRIN contract 20093BMNNPR.

\providecommand{\href}[2]{#2}\begingroup\raggedright\endgroup

\end{document}